\documentclass{IEEEtran}
\usepackage{amsmath,amssymb,amsfonts}
\usepackage{cases}
\usepackage{graphicx}
\usepackage{xcolor}
\usepackage{bbding}
\usepackage{multirow}

\IEEEoverridecommandlockouts                              

\title{\LARGE \bf
Learning the Dynamics of Future Marine Microgrids Using Temporal Convolutional Neural Network}

\author{Xiaoyu Ge$^{1}$, Ali Hosseinipour$^{2}$, Saskia Putri$^{3}$, Faegheh Moazeni$^{4}$, Javad Khazaei$^{5,\ast}$
\thanks{This work was in part under support from the Department of Defense, Office of Naval Research award number N00014-23-1-2602. The authors are with the Rossin college of engineering and applied science at Lehigh University, Bethlehem, PA 18015, USA. X. Ge, A. Hosseinipour, and J. Khazaei are with the Department of Electrical and Computer Engineering and S. Putri and F. Moazeni are with the Department of Civil and Environmental Engineering. (Emails: 
        {\tt\small xig620@lehigh.edu, alh621@lehigh.edu, sap322@lehigh.edu, moazeni@lehigh.edu,  khazaei@lehigh.edu)}}%
\thanks{$\ast$Corresponding author: Javad Khazaei. }
}

\begin{document}

\IEEEpubidadjcol
\IEEEpubid{\begin{minipage}{\textwidth}\ \\[25pt] \centering
\color{blue}This work has been submitted to the IEEE for possible publication. Copyright may be transferred without notice, after which this version may no longer be accessible.
\end{minipage}}

\maketitle

\begin{abstract}
Medium-voltage direct-current (MVDC) shipboard microgrids (SMGs) are the state-of-the-art architecture for onboard power distribution in navy. These systems are considered to be highly dynamic due to high penetration of power electronic converters and volatile load patterns such as pulsed-power load (PPL) and propulsion motors demand variation. Obtaining the dynamic model of an MVDC SMG is a challenging task due to the confidentiality of system components models and uncertainty in the dynamic models through time. In this paper, a dynamic identification framework based on a temporal convolutional neural network (TCN) is developed to learn the system dynamics from measurement data. Different kinds of testing  scenarios are implemented, and the testing results show that this approach achieves an exceptional performance and high generalization ability, thus holding substantial promise for development of advanced data-driven control strategies and stability prediction of the system.

\end{abstract}

\section{INTRODUCTION}
The onboard microgrid architecture on naval shipboards have been a subject of extensive research for decades. With the continuous advancement of technology, the naval shipboard microgrids (SMGs) have been growing both in size and complexity as a result of new propulsion motor technologies, advanced weaponry, energy storage systems, and even more complexity of power demand \cite{jin2016next,kumar2019comprehensive}. Conventionally, an alternative current (AC) architecture was implemented in SMGs, but inadequacies of an AC architecture has paved the way for direct current (DC) microgrids in SMG applications. Among the shortcomings of AC SMGs, one can allude to the fixed-speed operation of generators, which reduces efficiency, unwanted reactive power flow, imbalances and harmonics, and the need for voltage level conversion through bulky transformers~\cite{zheming}. On the other hand DC SMG benefit from simplified connection and disconnection of generation and storage devices, elimination of reactive power, reduction of the weight enables by variable high-speed generators, and elimination of phase angle synchronization~\cite{standard}. 

\par Despite offering many advantages as mentioned above, DC SMGs are highly volatile systems with fast dynamics, which can threaten the stability of these systems. Large and highly-dynamic load changes from pulsed loads, fast-switching power converters and their high-bandwidth control increase the probability of stability issues as a result of interaction dynamics and negative incremental impedance of constant power loads (CPLs)~\cite{uzair}. This highlights the importance of dynamic modeling for the purpose of stability assessment and control design. Conventionally, white-box models with containing differential algebraic equation were used for dynamic analysis of microgrids~\cite{multi}. However, due to confidentiality of models for many system components and uncertainty in the system model, system identification methods have gained popularity for data-driven model derivation of microgrids~\cite{identification}. 

\par Various system identification tools have been applied for identifying linear and nonlinear dynamical systems. The subspace method~\cite{subspace} and dynamic mode decomposition~\cite{dmd} methods are proposed for identifying linear dynamical systems, but cannot provide a good fit for the dynamics of nonlinear systems such as microgrids. Koopman operator can be applied for nonlinear system identification by approximating nonlinear equations in terms of linear functions. In~\cite{koopman}, Koopman operator is used for data-driven identification and predictive frequency control in power systems. Machine-learning based methods methods such as recurrent neural networks (RNNs)~\cite{lstm,rnn}, deep neural networks (DNNs)~\cite{deep,bansal2016learning}, convolutional neural networks (CNNs)~\cite{cnn}, deep Koopman methods~\cite{deep_koopman},  and ordinary differential equation networks (ODENet)~\cite{node} have also been applied for nonlinear system identification.  These methods utilize a neural network model that
is fitted to the system dynamics, relying on large training data sets and no prior physical knowledge of the system. Recently compressed sensing~\cite{compressed} and sparse regression~\cite{sindy, khazaei} techniques have been proposed to provide a trade-off between identification accuracy and model complexity of nonlinear systems. In~\cite{module}, a modularized method based on sparse regression method is developed to reduced the computational cost of predicting an AC microgrid dynamic model.
\begin{figure*}
    \centering
    \includegraphics[scale=0.6]{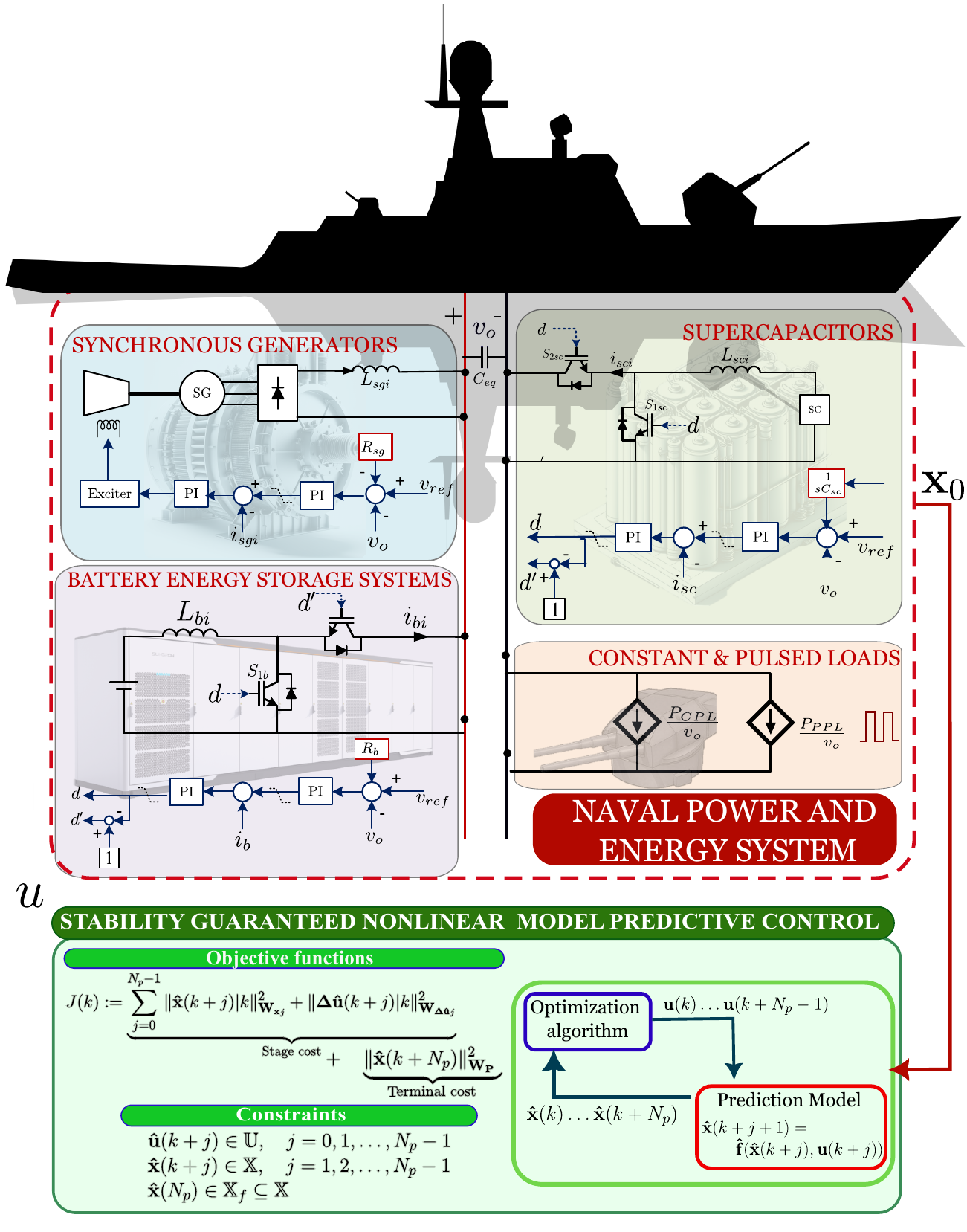}
    \caption{Topology of the MVDC SMG }
    \label{MVDC}
\end{figure*}
\par Despite their effectiveness, machine learning-based approaches such as DNN, RNN, and CNN require a large training data set, require future data for training (information leakage), and are computationally and memory intensive. Their practical applications are limited since they have to look into the past very far for more accurate predictions \cite{bai2018empirical}. In contrast to conventional machine learning approaches, temporal convolutional networks (TCNs) solve the above challenges by 1) employing causal convolution to prevent information leakage from the past to the present, 2) combining residual layers and dilated convolutions with very deep networks to look far into the history of data without requiring a lot of memory or computational power. Although TCN has been applied to time-series sequences in audio and signal processing applications \cite{hu2022hierarchic} and in power systems for battery health monitoring \cite{zhou2022battery} and load forecasting \cite{tang2022short} and showed superior performance compared with DNN and RNN, to the authors' best knowledge, no research has reported utilizing TCN for identifying dynamics of cyber-physical systems of a ship. Our objective in this study is to determine whether a data-driven model of a navy ship's power system that uses TCN can accurately identify asset dynamics. 

To address the above knowledge gap, dynamics learning of a hybrid medium-voltage DC (MVDC) SMG using a  temporal convolutional neural
network (TCN) framework is proposed. TCNs have been proven powerful in dealing with time-serious data prediction. The main contributions of the paper can be summarized as:
\begin{itemize}
    \item A TCN model is proposed to learn the transient dynamics of an MVDC SMG with supercapacitors, battery storage units, conventional generators. The proposed model can accurately identify the dynamics of the SMG in presence of pulsed power loads (PPLs) and constant power loads (CPLs).
    \item The proposed TCN can learn the complex power sharing using resistive and capacitive droop controllers among multiple energy resources that are designed to allocate supercapacitors for high-frequency load changes and batteries/generators for low-frequency state-state load changes.
\end{itemize}
In order to validate the effectiveness and generalization capacity of the proposed TCN-based data-driven modeling approach, several time-domain simulations in MATLAB were conducted and
and accurate results (up to 0.9999 $R^2$ value and 0.00754 MAE value) with a unseen dataset confirm the effective learning of the SMG dynamics with 100 seconds of training data at 0.5 milliseconds sampling rate. 
\section{MVDC Shipboard Microgrid}
A hybrid MVDC SMG with the topology shown in Fig.~\ref{MVDC} considered as the system under study. Two synchronous generators (SGs), battery energy storage systems (BESSs), and supercapacitors (SCs) are the sources of power that supply various kinds of loads in the SMG. Majority of high-power loads in a DC SMG such as propulsion motor controlled by speed drives and exhibit constant power load (CPL) behavior. 
The PPL represents the behavior of highly-dynamic mission loads in the SMG. 
\par The control diagram of each converter includes two cascaded proportional integral (PI) regulators to control the voltage and current of the converters for each resource. Detailed information on design of controllers can be found in \cite{hosseinipour2023multifunctional}. To enable power sharing between various resources, conventional droop control technique is implemented for SGs and BESSs to share steady-state power. SCs on the other hand are controlled using capacitive droop control method in order to act as a short-term energy storage to provide surge power compensation due to the high power density characteristics.  The control input for the BESS power converters are the duty cycles of their DC/DC converters, namely $d$ and $d'$. The control input for the SGs is the field voltage, denoted by $v_f$. Inner current and voltage controllers regulate the converter current and voltage, respectively. Due to the fact that dynamics of inner voltage and current controllers  are much faster than the droop controllers, the output voltage of SG and BESS units is decided by their droop controllers formulated by \cite{khazaei2021optimal}
\begin{align}\label{eq.droop}
  v_k=v_\text{ref}+\delta v-R_{k}i_{k}, ~\forall k \in\{\text{sga, sgb, ba, bb}\}
\end{align}
where $v_{\text{ref}}$ is the voltage setpoint, output currents of the SGs and BESSs are denoted by 
 $i_k$, and $R_k$ represent the droop gain for SGs and BESSs.  A secondary control input $\delta v$ is utilized to regulate the DC voltage of the main bus, $v_{bus}$, using a PI regulator with gains $k_{p}$ and $k_{i}$ formulated by
\begin{equation}
    \delta v=(k_{p}+\dfrac{k_{i}}{s})(v_{ref}-v_{bus})
\end{equation}
 
In order for the SCs to respond to transient power fluctuations, an integral droop control~\cite{integral_droop} decides their voltage reference as
\begin{align}\label{eq.droop}
  v_h=v_\text{ref}-\frac{1}{C_{h}}\int i_{h}, ~\forall h \in\{\text{sca, scb}\}
\end{align}
where $C_h$ is the virtual capacitance of the SCs, and $i_h$ denotes the output current of the SCs.

\begin{figure}[htp!]
    \centering
    \includegraphics[scale=0.75]{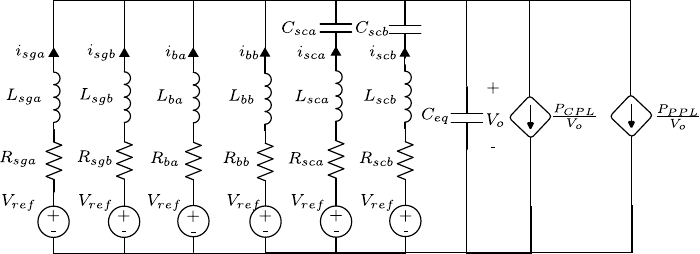}
    \caption{Reduced-order model of the MVDC SMG.}
    \label{MVDC_sim}
\end{figure}

 Owing to the high-bandwidth control of point-of-load converters including DC-DC converters and loads in Fig.~\ref{MVDC}, they can be represented by a lumped constant power load represented by the dependent current source~\cite{multiconverter}. The reduced-order model of the MVDC SMG in frequency domain taking into account droop control is given in Fig.~\ref{MVDC_sim}, where $s$ is the Laplace variable. The governing differential equations are resulted below after writing the node and loop equation in Fig.~\ref{MVDC_sim}.
\begin{align}{}
C_{\text{eq}} \frac{dV_o}{dt} &= \sum\limits_{i\in \mathcal{N}}i_{\text{i}} - \frac{P_{\text{CPL}}}{V_o} -\frac{P_{\text{PPL}}}{V_o} ~~\mathcal{N} =\{\text{sg,ba,sc}\}
\\\label{eqn:cap_voltage} 
    L_k\frac{di_{k}}{dt} &= V_{\text{ref}} - R_{k}i_{k} - V_o~~\forall k \in\{\text{sga,sgb,ba,bb}\} 
  \\\label{eqn:sga_current}
    L_{h}\frac{di_{h}}{dt} &= V_{\text{ref}} - R_{h}i_{h} - V_{h} - V_o ~~\forall h \in\{\text{sca,scb}\} \\\label{eqn:sca_current} 
C_{h}\frac{dV_{h}}{dt} &= i_{h} ~~\forall h \in\{\text{sca,scb}\}
\end{align}
\begin{figure*}
    \centering
    \includegraphics[scale=0.32]{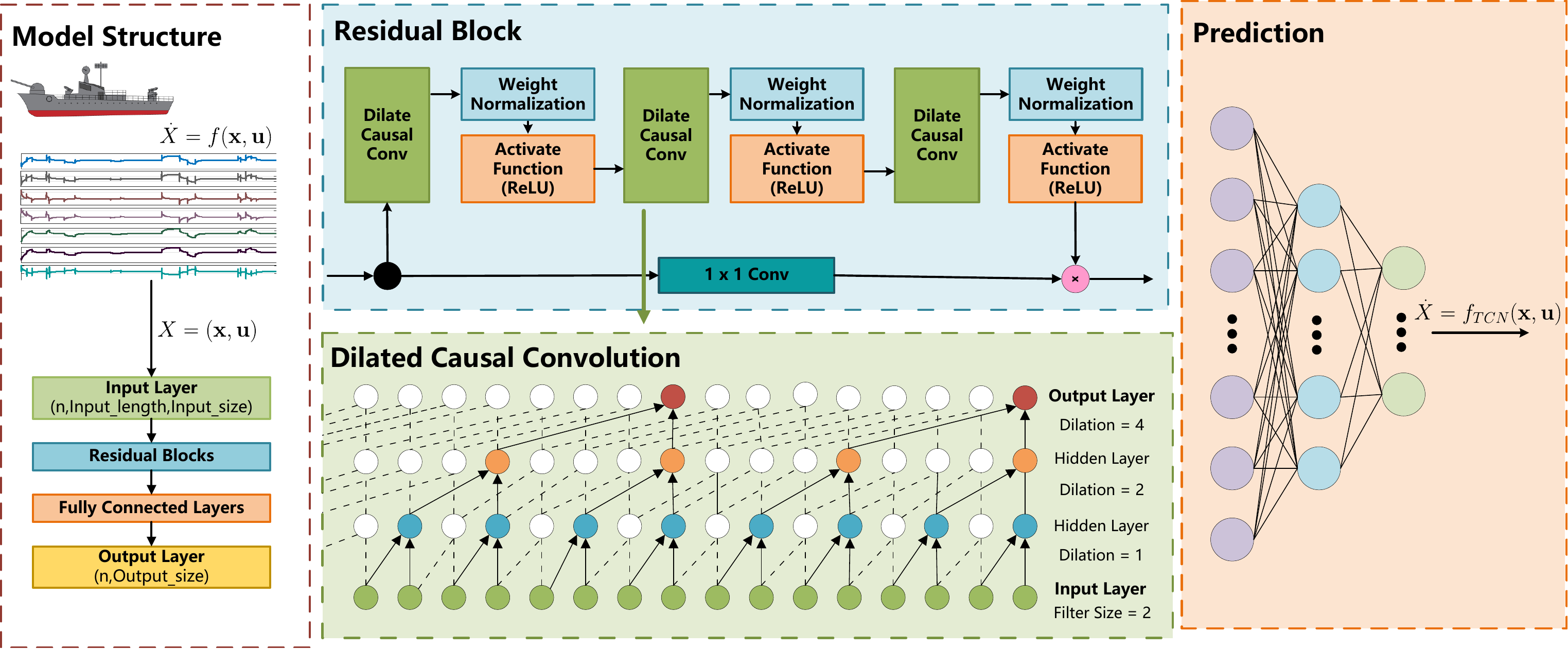}
    \caption{Structure of the proposed TCN architecture for identifying dynamics of SMGs.}
    \label{TCN}
\end{figure*}
\noindent \hspace{-0.1in} where, $V_{\text{ref}}$ is the bus voltage setpoint considered as the control input; $V_o$ is the MVDC voltage across the DC link capacitor $C_{\text{eq}}$; $P_{\text{CPL}}$, $P_{\text{PPL}}$ are the CPL and PPL power, respectively, representing the input disturbances; $i_{\text{sga}}$, $i_{\text{sgb}}$ are the output currents of the two SGs; $i_{\text{ba}}$, $i_{\text{bb}}$ are the output currents of the two BESSs; $i_{\text{sca}}$, $i_{\text{scb}}$ are the output currents of the two SCs; $R_{\text{sga}}$ and $R_{\text{sgb}}$ represent the SGs droop gains; $R_{\text{ba}}$ and $R_{\text{bb}}$ represent the BESSs droop gains; The capacitive droop gains of the SCs are also denoted by $C_{\text{sca}}$ and $C_{\text{scb}}$; The voltages across the virtual capacitors, $C_{\text{sca}}$ and $C_{\text{scb}}$, are signified by $V_{\text{sca}}$ and $V_{\text{scb}}$, respectively. Details on the complex droop control design for the ship and parameters of the system can be found in \cite{hosseinipour2023multifunctional}.

\section{Temporal Convolutional Neural Network}
The proposed TCN architecture for learning the dynamics of the SMGs is shown in Fig. \ref{TCN}. According to \eqref{eqn:cap_voltage}-\eqref{eqn:sca_current}, dynamics of the ship at time step $k$ can be represented by $\mathbf{\dot{x}}=f(\mathbf{x},\mathbf{u})$, where $\mathbf{x}$ is the vector of states and $\mathbf{u}$ is the vector of inputs. By collecting the state and input data over time and utilizing it for the training purposes, the proposed TCN structure utilizes residual blocks to learn the features of the data and a fully connected layer to predict the output ($\mathbf{\dot{x}}=f(\mathbf{x},\mathbf{u})$). The residual block within the TCN architecture includes causal convolutions with weight normalization and activation function as well as a fully convolutional network (FCN) as shown in Fig. \ref{TCN}.

\subsection{Dilated Causal Convolutions}
\begin{figure}
    \centering
    \includegraphics[scale=0.45]{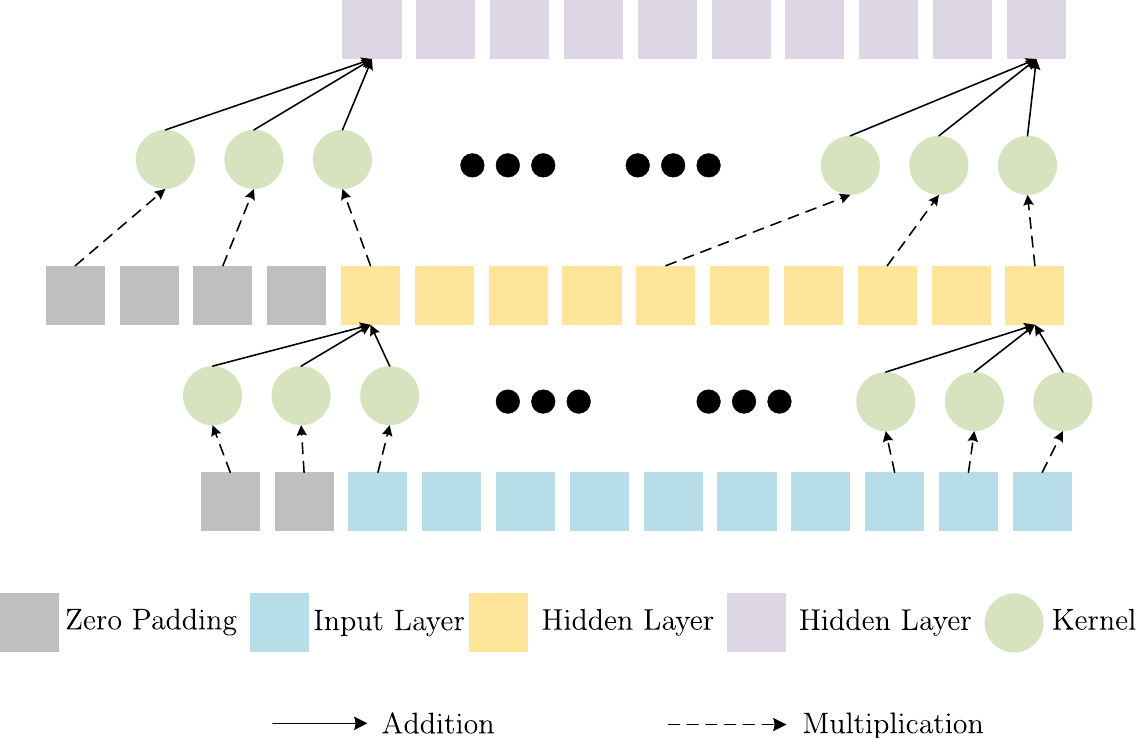}
    \caption{Dilated causal convolutions operation process.}
    \label{DCC_d}
\end{figure}

The causal convolution guarantees that there is no information leakage from the future, which means the $i \, \text{th}$ output in sequence $\{0,1,\dots, Input\_size-1\}$ only depends on the inputs with $\{0,\dots, i \}$ index. In this case, when the input history size is long, a deeper network and a larger filter is needed. In order to arrive at a more effective and feasible method, the dilated convolution is used here. In addition, the outputs should have the same length as the inputs, thus zero padding is added to the left part. The operation process is shown in \eqref{dcc_e} and Fig.\ref{DCC_d}.

\begin{equation}
    F_{DCC}(X,s)=\sum_{i=0}^{k-1}K(i)X_{s-di}
\label{dcc_e}
\end{equation}
where $\mathbf{X}=[\mathbf{x}, \mathbf{u}]$ is the matrix of collected data from the SMG, where $\mathbf{x}=[\mathbf{x}(t_0), \mathbf{x}(t_1),\hdots,\mathbf{x}(t_m)]$ collected state vector data for $t_m$ seconds, $\mathbf{x}=[x_1,x_2,\hdots,x_n]$ is the state vector, $K$ is the kernel, $k$ is the kernel size, $d$ is the dilation factor, and $s-di$ records the direction of operation \cite{bai2018empirical}. 

\subsection{Residual Connection} 
As models increase in depth, the gradients tend to diminish during the backpropagation process, leading to what is known as the vanishing gradient problem. This can severely hamper the effectiveness of model training. To address this issue, the inclusion of residual connections offers an auxiliary pathway for gradient flow, a technique that has consistently demonstrated its effectiveness in improving the training of very deep networks. This process is formulated by
\begin{equation}
    \mathcal{O}_{res}=F(\mathbf{X}+\mathcal{F}(\mathbf{X}))
\end{equation}
where $\mathcal{F}$ is the series of transformations and $F$ is the activation function.

\subsection{Weight Normalization}
In neural networks, the output of each layer is considered as the output of a nonlinear activation function whose input is the sum of weighted features, such as: 

\begin{equation}
    y=F(\mathbf{w}\cdot \mathbf{x}+b)
\end{equation}
where $\mathbf{x}$ is the input feature vector with a dimension of $k$, $\mathbf{w}$ is the weight vector with the same size as $x$, and $\mathbf{b}$ is the bias vector. During the training process, $\mathbf{w}$ and $\mathbf{b}$ will be updated with the back-propagation algorithm. To speed up the convergence of this process, the weight normalization is used after the 1-D fully convolutional layer~\cite{salimans2016weight}. The weight $\mathbf{w}$ can be normalized as :
\begin{equation}
    \mathbf{w}=\frac{g}{||\mathbf{v}||}\mathbf{v}
\end{equation}
where $g$ is a scaler, $\mathbf{v}$ is a $k$ dimensional vector and $||\mathbf{v}||$ is the Euclidean norm of $\mathbf{v}$. 

\subsection{Fully Connected Layer and Loss Function}
A fully connected layer is added after the residual block to make prediction based on the extracted features from the previous layers. In addition, to improve the generalization of the model, three fully connected layers are added here, the outputs size of the last fully connected layer is the same as the states we decided to predict. Throughout the training process, parameter optimization hinges on minimizing this loss function. Thus, the choice of the loss function significantly impacts model performance. Here, the MSE loss function, which is one of the most common ones in prediction tasks, is chosen. 
\begin{equation}
    \mathcal{L}oss_{MSE}=\dfrac{1}{N}||y_i-y_p||^2_2
\end{equation}
where $y_i$ is the ground truth value and $y_p$ is the prediction value.
\section{Case-Studies}
To show the accuracy of the proposed TCN-based dynamics learning method, several case studies are performed, including 1) the model capacity of generalization and 2) impact of training data size on the performance of the model. All models were run on a workstation with Intel Core i9-13900k CPU @5.8GHz, 32G RAM and NVIDIA GeForce RTX-4090 GPU.

\subsection{Datasets}
\begin{figure}
    \centering
    \includegraphics[scale=0.4]{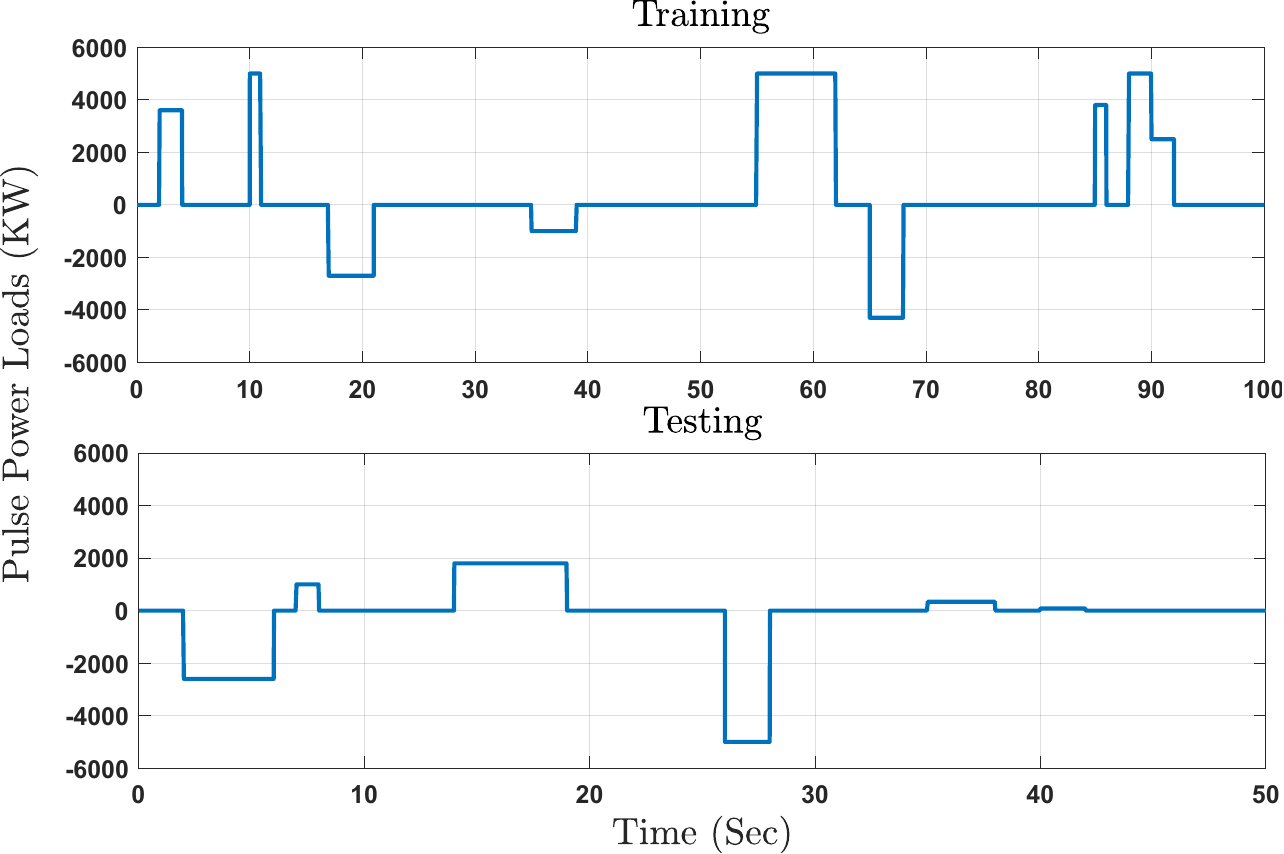}
    \caption{PPL variation for training and testing sets}
    \label{Pulse Data}
\end{figure}

\begin{figure}
    \centering
    \includegraphics[scale=0.42]{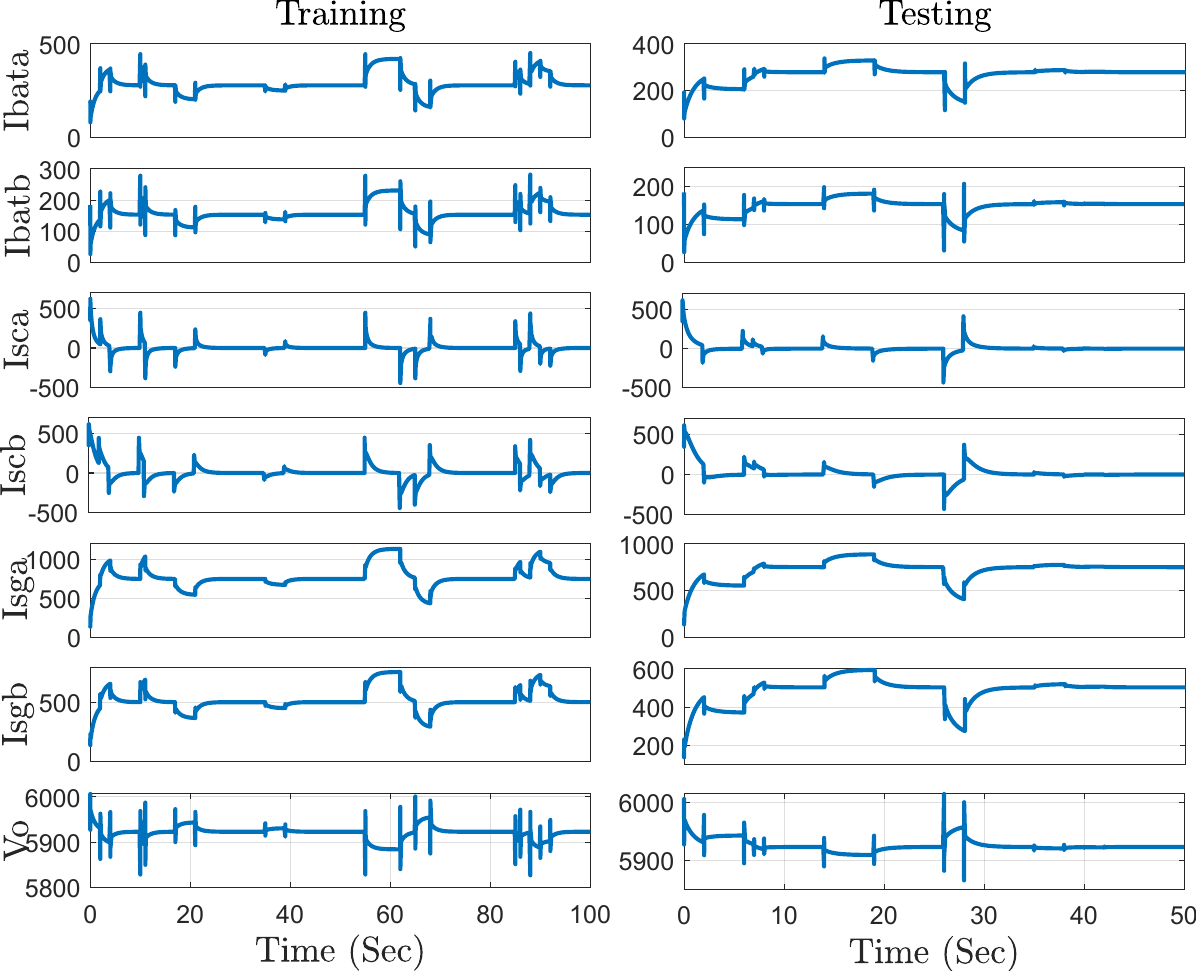}
    \caption{Training and testing datasets}
    \label{Train_test}
\end{figure}

The DC bus voltage, $V_o$, and the output currents of power sources, i.e., $i_{\text{sga}}$, $i_{\text{sgb}}$, $i_{\text{ba}}$, $i_{\text{bb}}$, $i_{\text{sca}}$, $i_{\text{scb}}$, are chosen as the target values for the proposed identification model, and  $P_{\text{PPL}}$ is the input disturbance collected for training and testing of the TCN. The parameters of the MVDC microgrid are given in Table~\ref{tab.simulation}. The PPL is varied from -5 to 5 MW with different pulse periods. The simulation is run for 100 seconds to collect the training data. To verify the performance of the TCN predictor, a new set of data with totally different PPL pattern is generated by running the simulation of the MVDC SMG for 50 seconds. The PPL variation is demonstrated in Fig.~\ref{Pulse Data}, and the corresponding state variables are shown in Fig.~\ref{Train_test}. After the data is collected, every 3000 time-steps data are seen as one train data, in other word, the history length of the each data is 3000.  

\subsection{Preprocessing of data}
Given the presence of multiple states, each exhibiting different values (e.g., $i_{sga}$ ranging from 100 to 200~A and $V_o$ fluctuating between 5800 and 6100~V), we have implemented min-max normalization. This approach ensures that all variables are normalized to a common scale, thereby enhancing computational efficiency.
\begin{equation}
    y=\frac{X-X_{min}}{X_{max}-X_{min}}
\end{equation}
where $X_{min}$ is the minimum value of the feature in dataset, and $X_{max}$ is the maximum value of the feature in dataset.
\subsection{Model Performance Evaluation}
To evaluate the model prediction accuracy comprehensively, the mean absolute error ($MAE$) and the R-squared ($R^2$) are computed. $MAE$ quantifies the average absolute difference between the ground truth and prediction value, while $R^2$ describes the fitting accuracy of the model.
\begin{equation}
    MAE=\frac{\sum_{i=1}^{N}|y_i-y_{pi}|}{N}
\end{equation}
\begin{equation}
    R^2=1-\frac{\sum_{i=1}^N{(y_i-y_{pi})^2}}{\sum_{i=1}^N{(y_i-y_m)^2}}
\end{equation}
where $y_i$ is the ground truth value, $y_{pi}$ is the prediction value, and $y_m$ is the mean value of the ground truth. 

\begin{figure}
    \centering
    \includegraphics[scale=0.44]{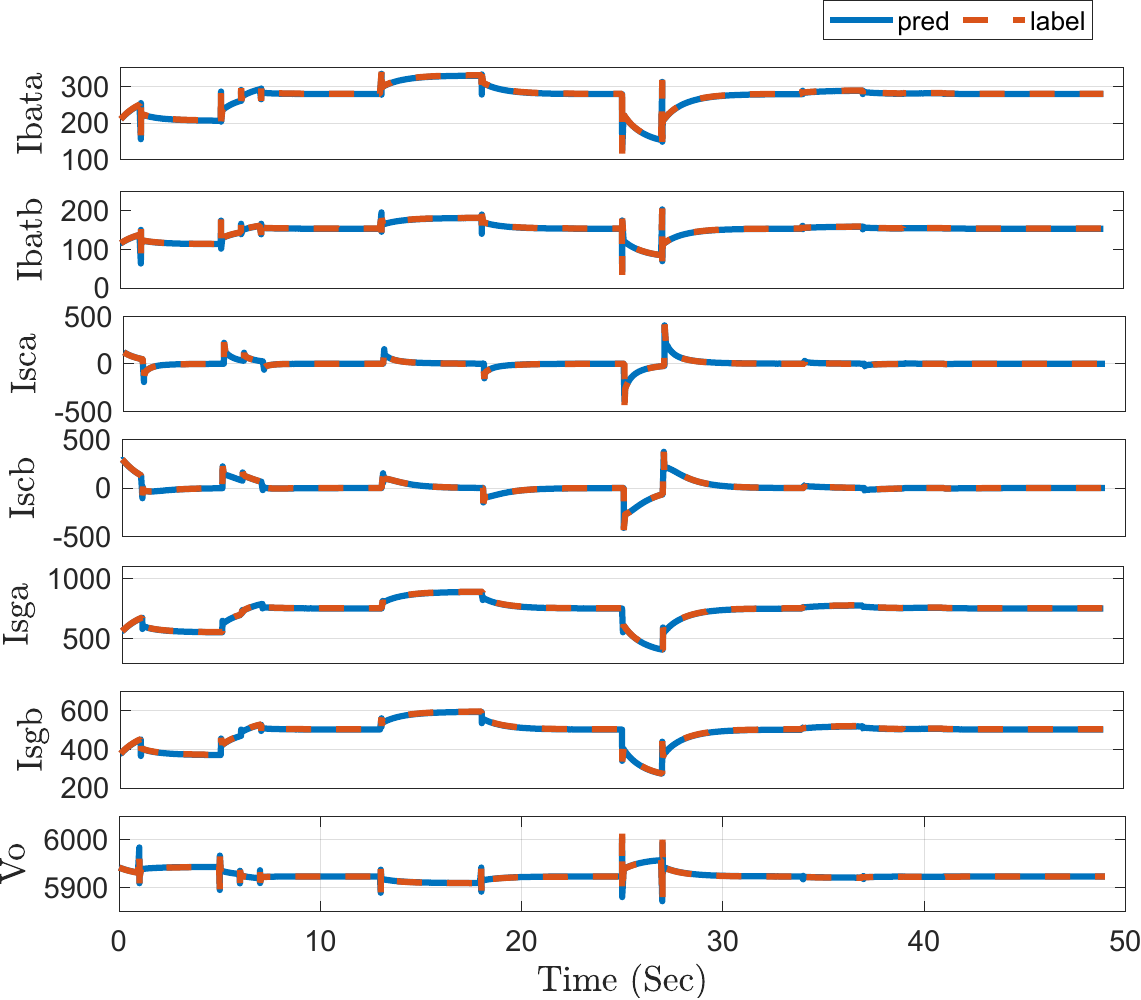}
    \caption{Dynamics learning performance}
    \label{results_n}
\end{figure}

The prediction results are shown in the Fig. \ref{results_n} and Table \ref{ner}. From both figure and numerical evaluation, it is clear that the proposed model exhibits high accuracy of prediction. Lower $MAE$ value and higher $R^2$ value imply that the model can capture the variation trend of the system states, and accurately predict all state variables for the next time step. 

\subsection{Generalization Potential of Model}
\begin{figure}
    \centering
    \includegraphics[scale=0.38]{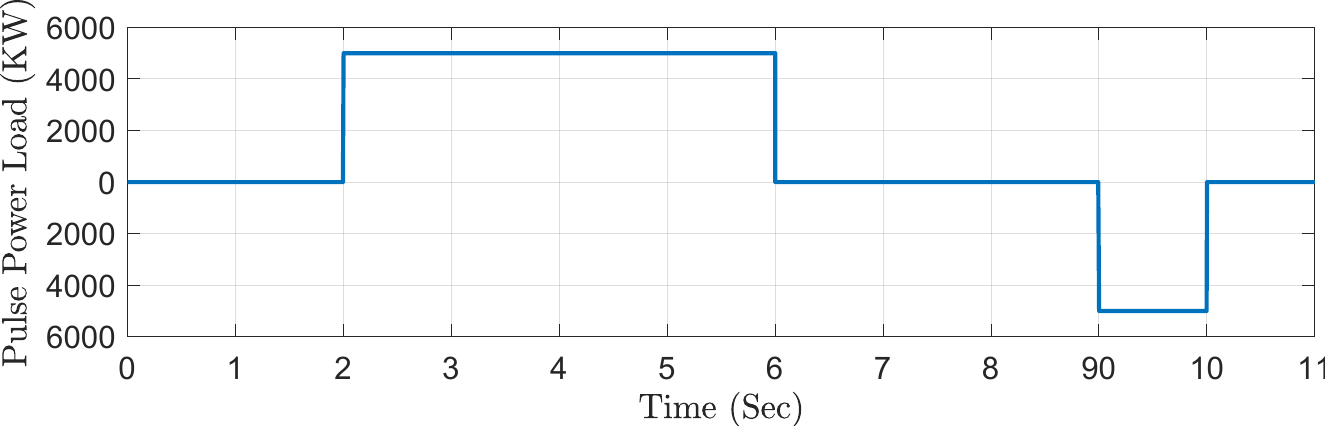}
    \caption{PPL variation within its minimum and maximum values.}
    \label{minmax}
\end{figure}

\begin{figure}
    \centering
    \includegraphics[scale=0.44]{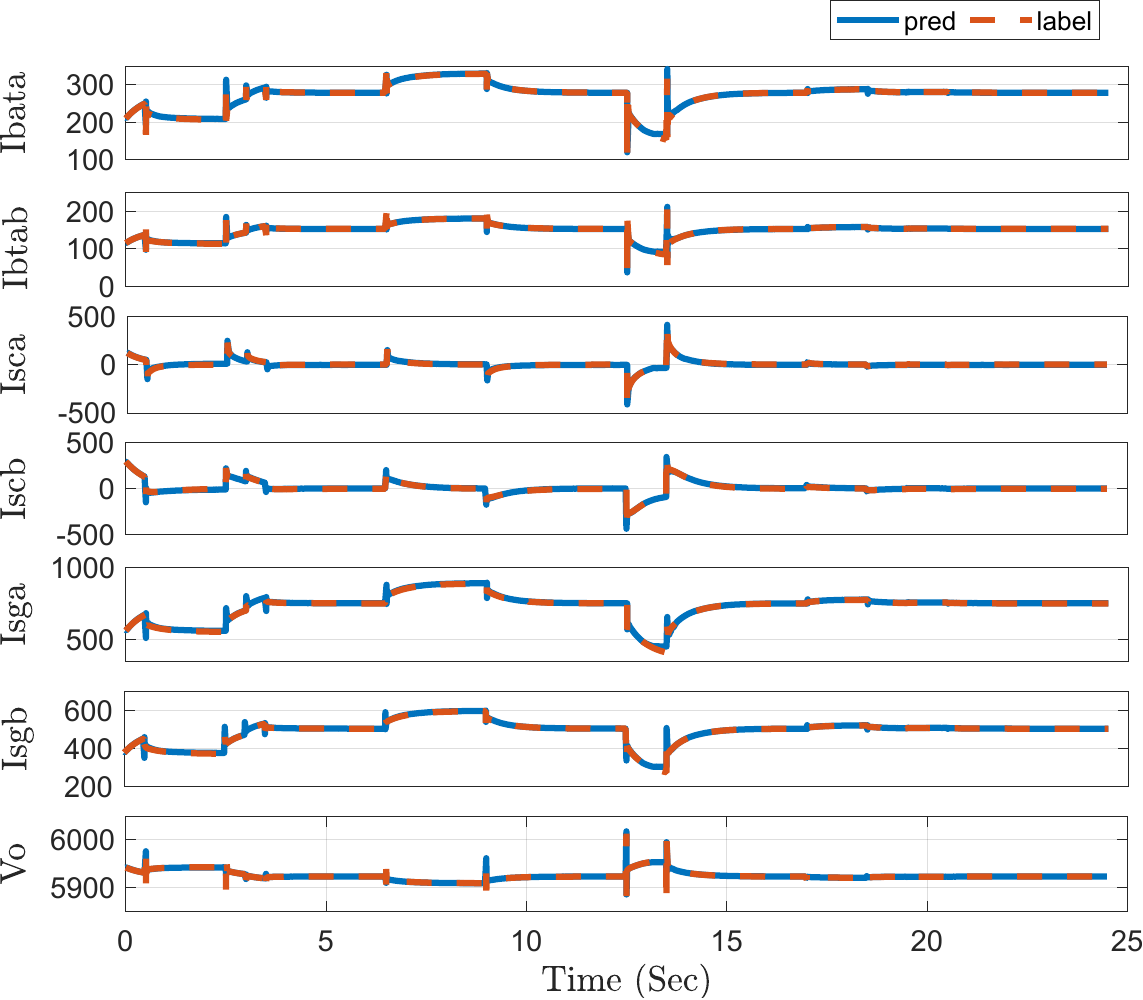}
    \caption{Testing results within Min-Max training set}
    \label{Res_minmax}
\end{figure}

The model generalization describes the the capacity of the model to deal with unseen data. To evaluate this, a new training dataset is generated by applying two pulses to the PPL with its amplitude changing between the maximum and minimum values with different duty cycles. The collected data is then used to retrain and test the model on the same testing set. The new PPL pattern used for training data generation is shown in Fig. \ref{minmax}, and the results of dynamics learning are shown in Fig. \ref{Res_minmax} and Table \ref{ner}. The results indicate that the model can accurately capture the dynamics of state variables and make precise predictions, whose average MAE is 1.7905, showcasing its robust generalization capabilities.

\subsection{The Impact of History Length}
This case study is designed to explore how the length of the history impacts the performance of the prediction model. Here we set the the history length to 1000, 2000, 3000 and 4000. The results of learning performance are shown in the Table.~\ref{ner}. From the table it can be seen that the model captures the variation trend very well with both history lengths ($R^2 \geq 0.99$). When the history length is changed from 1000 to 2000, the MAE increases by 42.48\%, and when the length increases from 2000 to 3000, the average MAE increases by 64.2\%. When the history length is 4000, the average MAE decreases by 69.6 \%.

\begin{table*}[]
\centering
\caption{Numerical Evaluation Results}
\label{ner}
\begin{tabular}{ccccccccccc}
\hline\hline
\multirow{2}{*}{} & \multicolumn{2}{c}{Length:1000} & \multicolumn{2}{c}{Length:2000} & \multicolumn{2}{c}{Length:3000} & \multicolumn{2}{c}{Length:4000} & \multicolumn{2}{c}{Min-Max} \\ \cline{2-11} 
                  & $R^2$           & MAE        & $R^2$           & MAE        & $R^2$           & MAE        & $R^2$           & MAE        & $R^2$            & MAE          \\ \hline
Ibata             & 0.9980      & 1.4109     & 0.9993      & 0.8114     & 0.9999      & 0.1776     & 0.9998      & 0.1045     & 0.9973       & 0.6270       \\
Ibatb             & 0.9911      & 1.6268     & 0.9988      & 0.5803     & 0.9998      & 0.1333     & 0.9996      & 0.1363     & 0.9966       & 0.4628       \\
Isca              & 0.9826      & 2.4676     & 0.9992      & 0.8197     & 0.9993      & 0.7928     & 0.9991      & 0.6698     & 0.9821       & 3.3138       \\
Iscb              & 0.9958      & 3.0030     & 0.9998      & 0.6302     & 0.9997      & 0.7995     & 0.9979      & 1.9527     & 0.9933       & 4.1838       \\
Isga              & 0.9994      & 1.6843     & 0.9994      & 2.0217     & 0.9999      & 0.1256     & 0.9999      & 0.4985     & 0.9969       & 2.5115       \\
Isgb              & 0.9996      & 0.6180     & 0.9993      & 1.5065     & 0.9999      & 0.0754     & 0.9999      & 0.2677     & 0.9973       & 1.1846       \\
Vo                & 0.9879      & 0.7547     & 0.9982      & 0.2827     & 0.9992      & 0.0774     & 0.9990      & 0.0768     & 0.9897       & 0.2498       \\
Avg               & 0.9935      & 1.6522     & 0.9992      & 0.9503     & 0.9997      & 0.3112     & 0.9993      & 0.5295     & 0.9933       & 1.7905       \\ \hline\hline
\end{tabular}
\end{table*}

\section{CONCLUSIONS}
In this study, a TCN model is proposed to learn the dynamics of MVDC shipboard microgrids. The simulation results indicate that the model can learn the system dynamics and predict the transient response of the system accurately. More importantly, the model shows high generalization capacity to deal with unseen data. All of those demonstrate our model has the potential to be used for control proposes, such as in data-driven model-predictive control (MPC). The future work may involve enhancing model generalization and accuracy through the incorporation of physical principles, as well as developing a controller to assess its closed-loop control performance.




 \section*{Appendix}
 Parameters of the SMG are shown in Table \ref{tab.simulation}.
\begin{table}
\renewcommand{\arraystretch}{1.3}
\caption{\footnotesize {Simulation parameters of the MVDC SMG.}}
\label{tab.simulation}
\centering
\scalebox{1}{
\begin{tabular}{c c}
\hline\hline
\bfseries {Parameter} & \bfseries {Value}\\
\hline
Simulation sample time & 50 $\mu$s\\
$R_\text{sga}$, $R_\text{sgb}$  & 0.05~$\Omega$, 0.1~$\Omega$\\
$R_\text{ba}$, $R_\text{bb}$  & 0.225~$\Omega$, 0.45~$\Omega$\\
$L_\text{sga}$, $L_\text{sgb}$  & 1~mH, 1~mH\\
$L_\text{ba}$, $L_\text{bb}$  & 0.8~mH, 0.8~mH\\
$L_\text{sca}$, $L_\text{scb}$  & 0.4~mH, 0.4~mH\\
$C_\text{sca}$, $C_\text{scb}$  & 5~F, 10~F\\
$C_{eq}$ & 10~mF\\ 
$P_{CPL}$ & 10~MW\\
\hline\hline
\end{tabular}}
\end{table}
    \section*{ACKNOWLEDGMENT}
 The authors would like to acknowledge the United States Department of Defense, Office of Naval Research for supporting this research.  

\bibliographystyle{IEEEtran}
\bibliography{IEEEabrv,root}

\end{document}